\documentclass[11pt,a4paper]{article}
\usepackage{times,latexsym}
\usepackage{url}
\usepackage[T1]{fontenc}
\usepackage{graphicx}
\usepackage{amsmath}
\usepackage{adjustbox}
\usepackage{multirow}
\usepackage{booktabs}
\usepackage{bbding}
\usepackage{array}
\usepackage{url}
\usepackage{xcolor}
\usepackage{soul}
\usepackage{colortbl}
\definecolor{lightgrey}{rgb}{0.9, 0.9, 0.9}
\sethlcolor{lightgrey}
\usepackage[acceptedWithA]{tacl2021v1}

\usepackage{xspace,mfirstuc,tabulary}

\newif\iftaclinstructions
\taclinstructionsfalse 
\iftaclinstructions
\renewcommand{\confidential}{}
\renewcommand{\anonsubtext}{(No author info supplied here, for consistency with
TACL-submission anonymization requirements)}
\newcommand{\instr}
\fi

\iftaclpubformat 

\else

\fi

\title{AIM: Let Any Multimodal Large Language Models Embrace Efficient In-Context Learning}

\author{
  Jun Gao
  \\
  \texttt{junegao1106@gmail.com}
  \And
  Ziqiang Cao
  \\
  \And
  Wenjie Li
  \\
}

\date{}

\begin{document}
\maketitle
\begin{abstract}
In-context learning (ICL) advances Large Language Models (LLMs) exhibiting emergent ability on downstream tasks without updating billions of parameters.
However, in the area of multimodal Large Language Models (MLLMs), two problems hinder the application of multimodal ICL:
(1) Most primary MLLMs are only trained on \textbf{single-image} datasets, making them unable to read extra multimodal demonstrations.
(2) With the demonstrations increasing, thousands of visual tokens highly challenge hardware and degrade ICL performance.
During preliminary explorations, we discovered that the inner LLM focuses more on the linguistic modality within multimodal demonstrations during generation.
Therefore, we propose a general and light-weighted framework \textbf{AIM} to tackle the mentioned problems through \textbf{A}ggregating \textbf{I}mage information of \textbf{M}ultimodal demonstrations to the latent space of the corresponding textual labels.
In image information aggregation, AIM independently generates \textbf{fused virtual tokens} to substitute each demonstration whose length is the same length as its texts.
Apart from significantly shortening length, fused virtual tokens modify the original multi-image prompts approximately to ones containing a single query image, effectively upgrading MLLMs pre-trained on single-image datasets to perform multimodal ICL.
We build AIM upon QWen-VL and LLaVA-Next, and we comprehensively evaluate AIM on image caption, VQA, and hateful speech detection.
Outstanding results reveal that AIM provides an efficient and effective solution in upgrading MLLMs for multimodal ICL.

\end{abstract}

\begin{figure}[h]
    \centering
    \includegraphics[width=.95\linewidth]{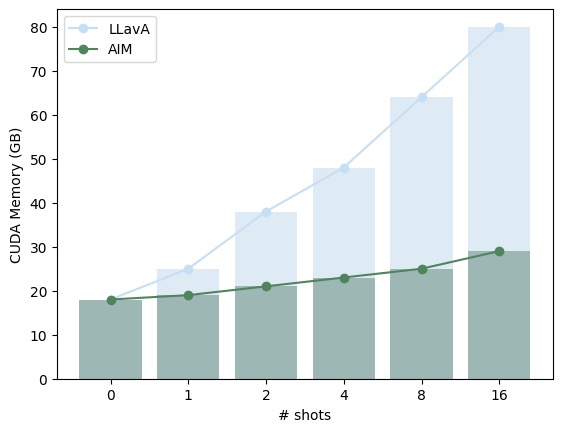}
    \caption{Memory cost comparison between AIM and LLaVA-Next on Flickr30k. The memory cost of LLaVA-Next occurs a surge, while it almost remains unchanged in AIM.}
    \label{fig:memory}
\end{figure}
\section{Introduction}
In-context learning (ICL) exhibits spectacular emergent ability in the NLP community~\cite{brown2020language,xie2021explanation,wang2023label}, enabling scaled-up Large Language Models (LLMs)~\cite{vaswani2017attention,wang2023chatgpt,yang2023exploring,wei2023zero,wang12023chatgpt,min2022noisy} to attain desirable performance on \textbf{training-agnostic} data by providing with handful in-context demonstrations.
Unfortunately, primary multimodal Large Language Models (MLLMs) such as LLaVA~\cite{liu2024visual,liu2023improved}, LLaMA-Adapter~\cite{zhang2023llama}, and BLIP-2~\cite{li2023blip}, only support a single image as the vision input, impossible to learn from multimodal demonstrations composed of \textbf{<image, instruction text, reference text>} pairs.
Additionally, most MLLMs utilize Perceiver~\cite{alayrac2022flamingo, awadalla2023openflamingo,liu2024visual,liu2023improved,liu2024llavanext,li2023blip,dai2024instructblip} to generate visual tokens from image features encoded by an existing visual encoder, assisting the inner LLM understand visual inputs.
However, multiple images in multimodal demonstrations inevitably produce thousands of visual tokens, resulting in extreme memory costs as depicted in Figure~\ref{fig:memory}.
Meanwhile, the prompt length surged by multimodal demonstrations might be one of the key factors constraining the performance of multimodal ICL~\cite{alayrac2022flamingo,awadalla2023openflamingo,QWen-VL,zhao2023mmicl,li2023mimic,huang2024language}.
Our experiments indicate a dramatic deterioration in the Perplexity (PPL) of answers generated by MLLMS with increasing demonstrations introduced (refer to Figure~\ref{fig:perplexity}).

\begin{figure}[t]
    \centering
    \includegraphics[width=\linewidth]{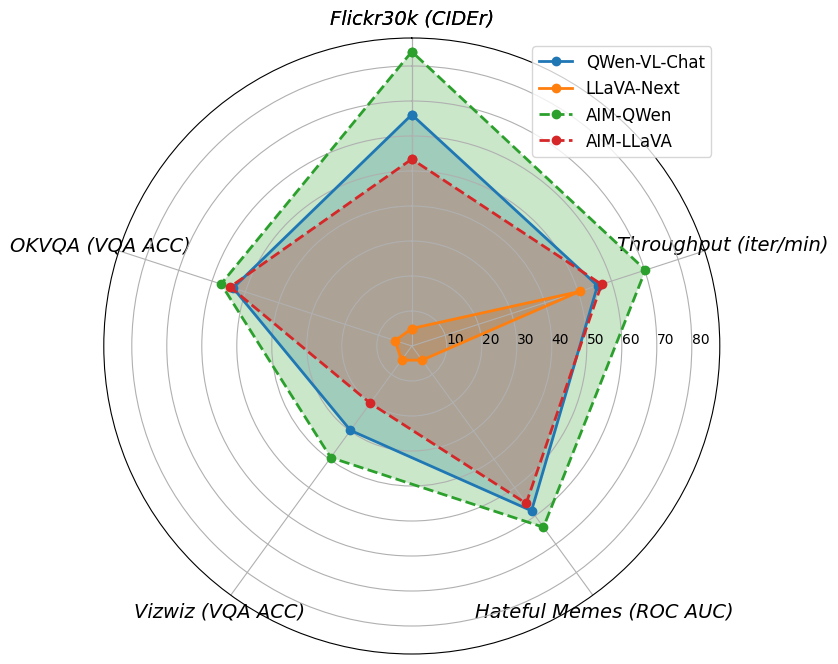}
    \caption{Performance comparison between AIM and its underlying backbone in the 16-shot ICL setting.}
    \label{fig:performance}
\end{figure}

\begin{figure}[t]
    \centering
    \includegraphics[width=\linewidth]{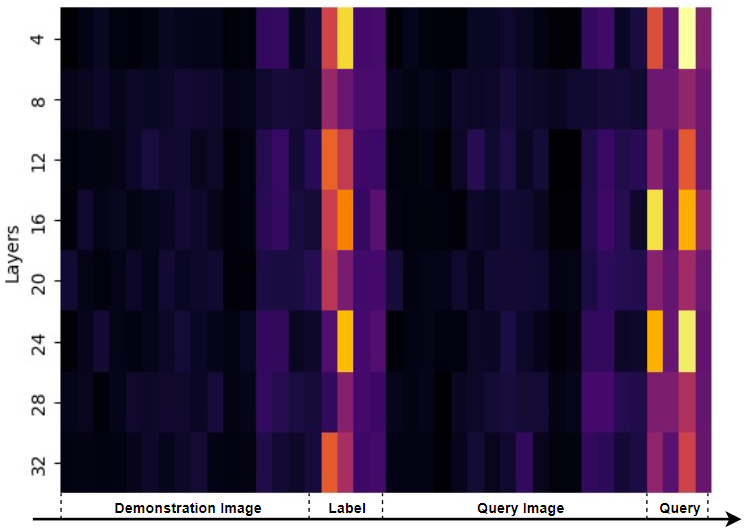}
    \caption{The hot map of attention scores when QWen-VL generates the first token on the hateful memes dataset. The brighter represents that responses to be generated have paid more attention to the current visual/textual tokens. Obviously, the generation relies more on the textual part of a multimodal demonstration}
    \label{fig:hotmap}
\end{figure}

During the early exploration, we surprisedly found that MLLMs attend more to the linguistic modality, namely the texts in demonstrations, than the visual modality for response generation, as shown in Figture~\ref{fig:hotmap}.
Motivated by this finding, we propose the framework AIM, with the \underline{aim} to make any MLLM embrace efficient multimodal ICL.
In contrast to mainstream MLLMs that always treat visual and textual tokens equally for both demonstrations and queries~\cite{QWen-VL,liu2023improved,liu2024visual,zhao2023mmicl}, AIM aggregates the image information of each demonstration into its linguistic modality and then drops their lengthy visual tokens.
Thus, AIM approximately reduces image-text demonstrations into text-like demonstrations with the images and texts in queries unmodified.
Specifically, AIM first applies the inner MLLM to forward each demonstration independently to obtain the hidden states of the image and its text.
Then, AIM applies a linear layer to project the hidden states on top of texts to fused virtual tokens acceptable for LLMs while dropping hidden states on top of images.
Therefore, the sizes of demonstrations are reduced to the dimensions of their textual embeddings.
Finally, the fused virtual tokens replace the original image-text pair, serving as a text-like demonstration with image information, fed into the inner LLM to guide response generation.
In this case, the built-in MLLMs are only required to attend to a single query image because images from demonstrations are removed in the input end.
Hence, AIM can perform ICL even if the backbones don't develop the understanding ability of interleaved multimodal inputs during pre-training.
Additionally, as aggregating image information is independent, AIM asynchronously processes different demonstrations within a batch to configure a virtual demonstration sequence for few-shot settings through horizontal concatenation.
The aggregated fused virtual tokens can be cached to formulate a demonstration bank (DB) for further reusing, avoiding repeated aggregation for the same demonstration.

AIM keeps its backbone frozen, only equipping 17M trainable parameters originating from the projection layer.
Considering ICL was proposed in a low-resource setting, previous studies training models on a mixture of downstream task datasets seem inappropriate that MLLMs perhaps fall into the short-cut answer, resulting in outstanding but not solid results on involved/related tasks. 
Hence, following the technical report of OpenFlamingo~\cite{awadalla2023openflamingo}, we train the projection layer on the subset of MMC4~\cite{zhu2023multimodal}, containing 223k images and 1M sequences from websites.
We select LLaVA-Next and QWen-VL as the underlying MLLM in AIM to verify the generality, representing MLLMs reading a single image-text pair and images interleaved with texts.
Furthermore, we comprehensively evaluate AIM on image caption~\cite{plummer2015flickr30k}, visual question answering (VQA)~\cite{gurari2018vizwiz,marino2019ok}, and hateful detection~\cite{kiela2020hateful}, none of the involved datasets occurring in the training data of AIM and mixture downstream pre-training dataset of QWen-VL and LLavA-Next.
Outstanding results in Figure~\ref{fig:performance} reveal that AIM always achieves comparable or better performance than its underlying backbone with less than 10\% tokens remaining on average (refer to Table~\ref{tab:ratio}).


Generally, our main contributions are as follows:
\begin{itemize}
    \item To the best of our knowledge, we are the first to analyze the attention distribution of multimodal demonstrations during generation, revealing that MLLMs prioritize attention towards the linguistic over the visual modality in multimodal demonstrations.
    \item Building upon this finding, we propose to transform multimodal demonstrations to text-like representations, enabling any MLLMs qualified for efficient multimodal ICL.
    \item Our proposed AIM exhibits efficiency in terms of trainable parameters, memory usage, and inference throughput.
\end{itemize}

\section{Related Work}
\subsection{Multimodal Large Language Models}
Recently, the development of LLMs significantly advanced the iterations of MLLMs, and the inner LLMs play crucial roles.
Researchers first trained the visual encoder to align to the frozen language models~\cite{tsimpoukelli2021multimodal}, performing vision-language tasks.
Predominate MLLMs can be abstracted to Perceiver \& LLM architecture, where the Perceiver is usually composed of a Vision Transformer (ViT)~\cite{dosovitskiy2020image} to extract image features and an adapter, concatenating visual and language tokens in the input end of the built-in LLM.
Specifically, the perceiver in QWen-VL~\cite{QWen-VL} and the Q-Former in BLIP-2~\cite{li2023blip} apply learnable queries to extract visual information based on cross-attention, while the Connector in LLaVA~\cite{liu2023improved,liu2024visual,liu2024llavanext} directly projects the visual features extracted from the pre-trained ViT-L/14 in CLIP~\cite{radford2021learning}.
Considering the further alignment within LLMs, Flamingo~\cite{alayrac2022flamingo} introduced the XATTN layer to align visual tokens originating from the Resampler and textual embeddings within the LLM.
InternLM~\cite{zhang2023internlm} propose Partial LoRA to align vision tokens to the LLM.

However, the perceiver in MLLM will introduce hundreds or even thousands of visual tokens to the inner LLM in ICL, resulting in over-length multimodal prompts and thereby bringing enormous memory costs.

\subsection{In-Context Learning}
In the field of NLP, LLMs including ChatGPT \cite{luo2023chatgpt}, GPT-4 \cite{openai2023gpt4}, and LLaMA \cite{meta2023introducing}, exhibit general spectacular emergent abilities on downstream tasks that provide a novel paradigm for generative models known as ``Pre-training, Prompting, and Prediction''.
Within this paradigm, ICL assumes a pivotal role, bolstering the generalization capability of LLMs \cite{wang2023chatgpt,yang2023exploring,wei2022chain}, without necessitating billions of parameters gradient updating.

The success of ICL in NLP boosts studies focusing on transforming it into the multimodal setting~\cite{tsimpoukelli2021multimodal,zhao2023mmicl,alayrac2022flamingo,yang2024exploring}.
Additionally, researchers extensively explore the influence of diverse demonstration configurations in captioning~\cite{yang2024exploring}.
As far as we know, recent studies focusing on multimodal ICL~\cite{alayrac2022flamingo,awadalla2023openflamingo,liu2023improved,zhao2023mmicl,li2023mimic} overlook deployment challenges to some extent.
QWen-VL~\cite{QWen-VL} and MMICL~\cite{zhao2023mmicl} treat visual and textual tokens equally during training, brought serious length challenges, and restricted model performance due to modeling enormous vision tokens.
Flamingo~\cite{alayrac2022flamingo} treated image information as informative noises adding to textual embeddings through extra introduced adapters within selectional inner LLM layers, resulting in additional module latency.
However, visual tokens of different images still share the same input window, bringing extra memory costs.
LLaVA-Next~\cite{liu2024llavanext} specialized in single-image inference, and it achieved outstanding performance on popular multimodal benchmarks and textual-only ICL, while its excellent performance failed to extrapolate to practical multimodal ICL settings, exhibiting poor ability of instruction following.
Additionally, LLaVA-Next connected pre-trained ViT and LLM via an MLP that resulted in thousands of visual tokens for high-resolution pictures, causing more serious length disasters than perceivers based on cross-attention.

\subsection{Efficient In-Context Demonstration}
Considering the huge inference costs brought by ICL, researchers recently focused more on formulating efficient in-context demonstrations~\cite{wingate2022prompt}.
Similar to prefix tunning~\cite{li2021prefix}, Gist Tokens~\cite{mu2023learning} were proposed to replace various task instructions.
AutoCompressor~\cite{chevalier2023adapting} first randomized segmented the texts with thousands of words into model-accepted range and then recursively generated soft prompts for each segment.
Similarly, ICAE~\cite{ge2023context} employed a LoRA-adopted Llama-7b \cite{touvron2023llama} to compress the processed demonstrations to compact virtual tokens.
\citet{gao2024selfcp,gao2024unifying} propose to compress over-limit prompts into virtual tokens via a frozen LLM and a linear layer.
Correspondingly, researchers also endeavored to shorten prompts by extracting informative tokens from the original ones \cite{li2023unlocking,jiang2023llmlingua}, namely token pruning \cite{kim2022learned} or token merging \cite{bolya2022token}.
LLMLingua \cite{jiang2023llmlingua} and Selective Context \cite{li2023unlocking} shared similarities but diverged on whether to eliminate tokens with high or low PPL.

Unfortunately, these outstanding studies only focus on the textual modality, which did not suffer from the modal gap in MLLMs.
To our best known, AIM is the first to explore the construction of efficient multimodal demonstrations.

\begin{figure*}[t]
    \centering
    \includegraphics[width=.95\textwidth]{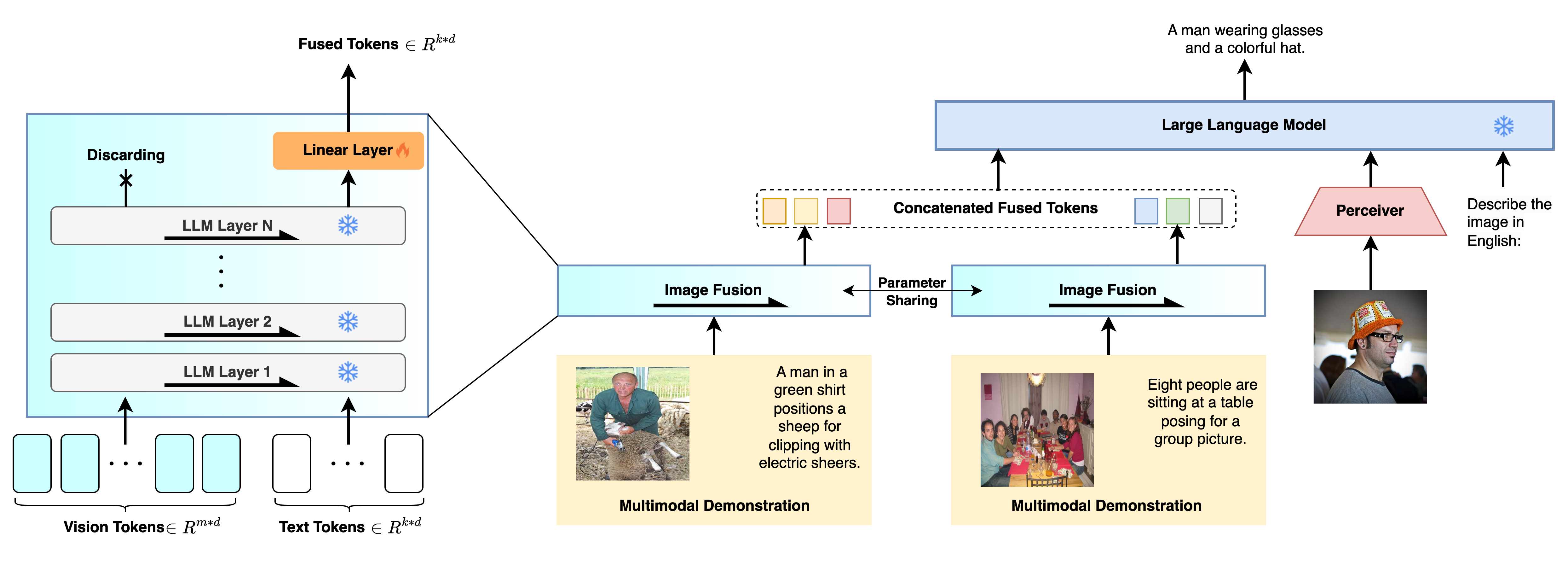}
    \caption{The architecture of AIM. Fused tokens from different demonstrations are concatenated and fed into the inner LLM, discarding original visual tokens.}
    \label{fig:model}
\end{figure*}

\section{Methodology}
\begin{table}
\centering
\begin{adjustbox}{width={\linewidth},totalheight={\textheight},keepaspectratio}
\begin{tabular}{l|cccccc}
\bottomrule
Methods  & In-context Learning Mode & Inner LLM & \# Trainable Para. \\
\hline
Flamingo & multimodal & Chinchilla (7B) & 7B  \\

MMICL & multimodal & FLANT5 (3B/11B) &   $\gg$17M   \\

QWen-VL & multimodal & QWen (7B)&   7B  \\

LLaVA-Next & Language-only & Vicuna (7B) & 7B  \\

\hline
AIM & multimodal & QWen/Vicuna (7B)  & 17M \\

\toprule
\end{tabular}
\end{adjustbox}
\caption{Quality comparison of recent MLLMs and AIM in ICL mode, LLM size, and trainable parameters.}
\label{tab:comparision}
\end{table}
We propose AIM as illustrated in Figure~\ref{fig:model}, which is a training- and inference-efficient framework that aggregates image information of multimodal demonstrations into their latent space of texts.
Considering different details of popular MLLMs, we present an empirical comparison in Table~\ref{tab:comparision}.
Specifically, The Flamingo is distinguished from others by its Gated XATTN layer inserted in the LLM blocks to fuse image information into embeddings, \textbf{sacrificing inference efficiency to memory usage}.
LLaVA-Next directly projects visual features extracted from pre-trained ViT to thousands of visual tokens, \textbf{resulting in higher memory cost increment} than other methods based on Q-Former.
Additionally, LLaVA-Next can read only a single image and thus it doesn't support multimodal ICL.

Similar to Flamingo, AIM discards substantial visual tokens in multimodal demonstrations after aggregating demonstrated image information into its text, resembling a multimodal ICL prompt approximately containing a single query image.
Therefore, AIM operates seamlessly with any MLLMs, regardless of whether they initially understand multimodal demonstrations well.
AIM employs the 7B version of QWen-VL and LLaVA-Next as the built-in backbone, representing the two coarse-grained types of MLLMs divided by input form, to verify the effectiveness of fused virtual tokens.

\subsection{Preliminary}
A multimodal ICL prompt encompasses several demonstrations consisting of image-text pairs $(<X_1, Y_1>, <X_2, Y_2>, ..., <X_n, Y_n>)$ and a query denoted as $<X_{query}, ins.>$, where $X_i$ and $X_{query}$ representing the $i$-th demonstration image and the query image.
Additionally, we use manually designed instructions $ins.$ to wrap the bare label for each demonstration.
Thus, the demonstrated texts in $i-$th demonstration are composed of instruction $ins.$ and its reference label $Y_i$, e.g., \textbf{[IMG]} Describe the image in a sentence in English: \textbf{[Caption]}. 

\subsection{Image Information Aggregation}
Taking into account our discoveries, AIM configures efficient demonstrations by disrupting the original parity between visual and linguistic modalities.
AIM signals the linguistic space to gather image information via the forward propagation of the inner LLM, as illustrated in the left part of Figure~\ref{fig:model}.
Practically, we split the original concatenated $n$-shot demonstrations into $n$ separate image-text pairs, decorating labels with manually designed instructions.
Then, we feed images to the Perceiver, the Visual Prompt Generator (VPG), to obtain the visual tokens $(X^v_1, X^v_2, ..., X^v_n)$ normally.
Consequently, the first $N$ LLM layers forwards $X^v_i$ attached with the $i$-th textual embeddings $Y_i$, obtaining last hidden states corresponding to $Y_i$, while dropping the others\footnote{$\oplus$ means token-level concatenation.}:
\begin{equation}
   \_, H^Y_i =\text{Forward}(X^v_i \oplus Y_i).
\end{equation}
Due to the inner transformer layers, $H^Y_i$ is compelled to attend to the preceding visual tokens, making the latter textual tokens able to aggregate visual information.
However, $H^Y_i$ is still in the output space that the LLM can't understand directly although it has fused with image information.
We insert a learnable projection layer serving as the adapter to convert $H^Y_i$ into the LLM-acceptable fused tokens, similar to the perceiver converting visual features from visual encoder to visual tokens:
\begin{equation}
    \hat{Y_i} = W_p\cdot H^Y_i,
\end{equation}
where $W_p$ is the parameters of the projection layer.
Notably, aggregating image information of each image is independent of other demonstrations.
Thus, AIM supports obtaining all $\hat{Y_i}$ in a batch asynchronously or repeating this process for each demonstration synchronously to trade off memory costs with time.
\begin{table}
  \centering
    \begin{adjustbox}{width={\linewidth},totalheight={\textheight},keepaspectratio}
    \begin{tabular}{l|c|c|c|c|c}
    \toprule
    Dataset & Training &\# Instances & Eval. Set &  Eval. & Metric\\
    \hline
    Flickr30k & \multirow{4}{*}{\XSolidBrush} & 1000 & Test(Karpathy)& Open-Ended & CIDEr \\
    OKVQA &  & 5046 & Val & Open-Ended & VQA acc.\\
    Vizwiz &  & 4319 & Val & Open-Ended & VQA acc.\\
    Hateful Memes &  & 815& Seen Test & Cloese-Ended & ROC AUC \\
    \hline
    \end{tabular}%
\end{adjustbox}
    \caption{Details of involved evaluating datasets.}
  \label{tab:dataset}%
\end{table}%

\subsection{Response Generation in multimodal ICL}
AIM applies the entire frozen inner LLM to respond to current queries, with the guidance of our proposed fused tokens.
For $n$-shot ICL, AIM obtains $(\hat{Y_1}, \hat{Y_2}, ..., \hat{Y_n})$ independent from each other, concatenating them to configure an efficient demonstration sequence $D=\hat{Y_1}\oplus\hat{Y_2}\oplus...\oplus\hat{Y_n}$.
Finally, the demonstration sequence $D$ together with query image $X_{query}$ and the instruction $ins.$ are fed into the inner LLM, performing auto-regressive generation:
\begin{equation}
    y_t = argmaxP(y|D; X_{query}; ins. ; y_{<t}).
\end{equation}


\subsection{Training}
The trainable parameters in AIM are merely 17M originating from the projection layer $W_p$.
We supervised-tune the projection layer under the language modeling objective of its built-in LLM.
We collect 56k instances from the web multi-image dataset MMC4.
Each instance includes several images $[X_1, X_2, ..., X_k]$, and each image corresponds to a most similar text $[Y_1, Y_2, ..., Y_k]$ assigned by existing ViT-L/14 in CLIP, constructing an interleaved image-text training instance.
During data preprocessing, we ensure each training instance has non-overlapping remaining texts and concatenate them, denoted as $Y^R = Y_{k+1} \oplus Y_{k+2} \oplus ...$.

AIM first independently aggregates $X_i$ to its corresponding text $Y_i$, obtaining $\hat{Y_i}$.
Then, the language modeling loss can be formulated as:
\begin{equation}
    loss = -\frac{1}{|Y^R|}\sum_{t=0}^{|Y|}logP(Y^R_t|\hat{Y_1}, ..., \hat{Y_k}; Y^R_{<t}).
\end{equation}

Notably, the carefully designed training approach separates the information aggregation of different images, breaking the inner relevance among images crawled from the same web page, increasing learning difficulty, and scaling up model robustness.
Specifically, this gets rid of the influence of other image-text pairs, allowing each image to focus on the given text only and better match the patterns of ICL intuitively.
More importantly, assuming $k$ image-text pairs of length $l$, breaking their inner relevance will reduce the memory complexity from $\mathcal{O}(k^2l^2)$ to $\mathcal{O}(kl^2)$.
It also brings AIM crucial merit that each aggregated result can be cached independently, formulating a demonstration bank (DB) for further reusing without aggregating image information into its latent space of demonstrated texts every time.

\section{Experiment}
\subsection{Dataset}
We briefly illustrate the involved dataset of AIM in Table~\ref{tab:dataset}.
Involved evaluating datasets are filtered according to the mixture downstream training dataset of the underlying MLLMs to simulate the ICL scenario.
\subsubsection{Training Data}
\paragraph{multimodal C4} MMC4 is a public open, billion-scale corpus of images interleaved with text.
CLIP ViT-L/14 assigns each image with the most matched text crawled from the web.
Due to AIM being trained parameter-efficiently, we merely utilize a subset of 56k instances to formulate our training set, which contains 223k images and 1M texts.
\subsubsection{Evaluating Data}
\paragraph{Flickr30k}
Flickr30k is a popular captioning dataset that contains 31,000 images collected from Flickr with 5 references annotated by humans.
\paragraph{OKVQA} OKVQA, Outside Knowledge Visual Question Answering, includes over 14k manually filtered questions that require external knowledge, with 5 ground truth answers for each question respectively.
\paragraph{Vizwiz}
Vizwiz~\cite{gurari2018vizwiz} dataset originates from a natural VQA setting where blind people each took an image and recorded a spoken question about it, together with 10 crowdsourced answers for each image.
Especially, Vizwiz requires model responding ``unanswerable'' when the provided image is insufficient to answer the question.
\paragraph{Hateful Memes}
Hateful Memes is a binary classification dataset to detect hateful speech in multimodal memes.
Their mentioned ``Seen/Unseen'' Test sets are distinguished by whether it is described in their paper or their challenge competition report.

\subsection{Evaluation Metrics}
We widely evaluate AIM on a spectrum of vision-language benchmarks, including Flickr30k, OKVQA, Vizwiz, and Hateful Memes.
Following previous studies, we use CIDEr for Flickr30k, VQA-ACC for OKVQA and Vizwiz, and ROC AUC scores for Hateful Memes.
All the evaluation scripts are publicly available on the repository of OpenFlamingo.
\begin{table*}
\centering
\begin{adjustbox}{width={\textwidth},totalheight={\textheight},keepaspectratio}
\begin{tabular}{l|c|cccccc|cccccc|cccccc|cccccc}
\bottomrule
\multirow{4}{*}{Method} & \multirow{4}{*}{LLM}  & \multicolumn{6}{c|}{Flickr30k}  & \multicolumn{6}{c|}{OKVQA} & \multicolumn{6}{c|}{Vizwiz} & \multicolumn{6}{c}{Hateful Memes} \\
& & \multicolumn{6}{c|}{\textit{\small CIDEr $\uparrow$ }}& \multicolumn{6}{c|}{\textit{\small VQA-ACC $\uparrow$}}& \multicolumn{6}{c|}{\textit{\small VQA-ACC $\uparrow$}}& \multicolumn{6}{c}{\textit{\small ROC-AUC $\uparrow$}}\\
\cline{3-26}
& & \multicolumn{6}{c|}{\textit{\small\#-shots}}& \multicolumn{6}{c|}{\textit{\small\#-shots}}& \multicolumn{6}{c|}{\textit{\small\#-shots}}& \multicolumn{6}{c}{\textit{\small\#-shots}}\\
&& 0& 1 & 2  & 4& 8& 16 & 0& 1 & 2  & 4& 8& 16 &  0& 1 & 2  & 4& 8& 16 &  0& 1 & 2  & 4& 8& 16 \\
\hline
\textit{Flamingo}$^{\dagger\bowtie}$  & Chinchilla (7B)  & 61.5 & - & -& 72.6& -& - & 44.7 & - & -& 49.3& -& - & 28.8 & - & -& 34.9& -& - & 57.0 & - & -& 62.7& -& - \\
\hline
\textit{Open Flamingo}$^{\dagger\bowtie}$   & \multirow{2}{*}{MPT (7B)}  
& 39.2 & - & -& 52.2& 58.7 & 60.6 & 
38.3 & - & -& 42.0 & 44.1& 45.1 &
34.6 & -&- & 41.0 & 45.0& 46.2& 
67.1 & - & -& 70.1 & 71.2& 73.2 \\
\textit{\quad\quad\quad-Random}$^{\dagger\bowtie}$  &   
& 59.5 & - & -& 65.8& 62.9 & 62.8 & 
37.8 & - & -& 40.1 & 41.1& 42.7 &
27.5 & - & -& 34.1 & 38.5& 42.5& 
51.6 & - & -& 54.0 & 54.7& 53.9 \\
\hline

\textit{QWen-VL $^\star$} & QWen (7B)
&{73.4}&74.6&75.3&77.1&75.1&63.0 & 
{46.3}&42.9&51.2&52.6& 53.2& 53.3 & 
{27.6}&28.3	&29.2&30.6& 30.1& 28.7&
{56.5}&58.0&57.3&56.2& 57.1& 59.5  \\
\hline

\textit{QWen-VL} & \multirow{5}{*}{QWen (7B)} 
&\multirow{5}{*}{74.3}&\textbf{75.6}&79.0&80.4&74.6&66.1 & \multirow{5}{*}{55.3}&45.5&55.9&56.4& 55.1& 53.7 & \multirow{5}{*}{28.4}&29.5	&30.9&31.3& 31.7& 29.8 &\multirow{5}{*}{58.3}&57.8&59.1&59.0& 59.3& 58.2  \\

\textit{\quad\quad-w/o visual} & &
&74.5 & 75.1 & 74.7 & 74.3 & 71.7& 
& 53.2 & 54.0 & 54.7 & 55.3&  55.1&
& 29.1 & 30.3& 28.7 & 30.5& 29.8&
& {58.0}& 57.7& 58.2& 58.2 & 57.1  \\

\textit{AIM} &  & 
&67.6&76.2&78.1&78.8&82.3 &
&55.6&55.4&55.8&56.1&56.0 &
&\textbf{34.2}&34.3&35.1&35.6 & 36.1 &
&\textbf{58.0}&\textbf{60.0}&57.8 & 58.9 & 57.1 \\

\quad\quad\textit{-16}  & &
&55.1&\textbf{79.7}&\textbf{81.2}&\textbf{83.4}&\textbf{84.1}&
&\textbf{57.3}&\textbf{57.3}&\textbf{57.9}& \textbf{58.2}&\textbf{57.3} &
&30.9&\textbf{35.9}&\textbf{37.6}&\textbf{38.3} & \textbf{39.4}&
&55.6&56.4&\textbf{59.6}& \textbf{62.9}& \textbf{64.0}  \\

\quad\quad\textit{-24} &  &
&63.9&74.4&73.3&75.8&78.1 & 

&56.2&56.6&55.2&54.3& 53.5 & 
&31.8&34.2&34.3&34.6& 34.6 &
&52.5&57.1&56.9& 57.4& 59.1  \\
\hline

\textit{LLaVA-Next $^\star$} &Vicuna (7B)
&44.6&38.5&32.7&23.5&<1& <1 & 
{45.3}&30.1&33.5&29.8& 28.6 & <1 & 
{18.7}& 16.7 & 17.8 & 14.5& 15.9& <1 &
{55.3}& 54.6& 54.0& 52.8& 52.4 & NaN\\
\hline

LLaVA-Next & \multirow{5}{*}{Vicuna (7B)}  & \multirow{5}{*}{46.2} & <1 & <1& <1& <1& <1 & \multirow{5}{*}{49.7} & <1 & <1& <1& <1& <1 & \multirow{5}{*}{19.8}&16.4&14.0&11.2& <1& <1 &\multirow{5}{*}{55.9}&53.6&54.7&55.6& NaN& NaN  \\
\textit{\quad\quad-w/o visual} & 
&& 47.2& 47.7& 48.1 & 47.6 & 44.3 & & 51.2 & 52.6 & 52.9 & 51.5& 52.8& & 21.1 & 20.6 & 21.7& 20. 5& 19.9 &&{56.1}& 56.5& 55.9& 55.6&  55.0\\
\textit{AIM} & & &50.1&54.5&58.2&\textbf{57.5}&\textbf{53.4}&&\textbf{54.0}&\textbf{55.1}&\textbf{55.3}&\textbf{55.0}&\textbf{54.6}& &23.1&23.8&\textbf{24.1}&\textbf{21.5}&\textbf{20.2}&&55.6&\textbf{56.6}&\textbf{58.6}&\textbf{57.8}&\textbf{55.6}\\
\quad\quad\textit{-16} &    & &49.5&51.7&44.3&30.8&26.5&&51.8&51.8&51.7&50.6&48.7 & &21.4&20.9&21.8&17.4&14.8&&55.5&55.7&56.3&53.4&53.0 \\
\quad\quad\textit{-24} &   & &47.1&47.5&46.0&39.2&38.0&  & 53.3&54.7&54.3&\textbf{55.0}&51.4& &\textbf{23.3}&\textbf{23.8}&21.6&20.0&19.0&&\textbf{55.7}&56.4&55.3&54.7&53.3\\

\toprule
\end{tabular}
\end{adjustbox}
\caption{Main results of AIM. \textit{w/o visual} stands for providing textual label only. -16/24 represents the number of LLM layers applied to aggregate image information. $\dagger$ stands for the results from previous works and $\bowtie$ indicates extra providing 2 textual label in 0-shot. $\star$ represents further tune backbones with LoRA. \textbf{The 0-shot results of AIM and its backbone are the same and merged because no demonstrations are provided for aggregation.}}
\label{tab:results}
\end{table*}

\subsection{Setting}
During training, we set the maximum number of pictures to 5 per step for efficiency and filtered images if their similarities with all texts below 0.24 following OpenFlamingo.
We fix the learning rate to 3e-5 and use Adam as the optimizer, and the effective batch size is 16 (4 GPUs data parallelism and 4 steps gradient accumulation).
The number of epochs is set to 10 and we get a checkpoint per 3400 training steps.
Additionally, we conduct all experiments on a single DGX node with 8*Nvidia H800 GPUs.
LLaVA-Next supports processing any resolution image by splitting it into sub-images, bringing several times visual tokens.
We ignore this character and require LLaVA to pad each image to 336*336 resolution since AIM introduces mass pictures as demonstrations\footnote{Set image\_aspect\_ratio to pad.}.

We borrow some crafted prompts from previous studies.
For captioning, we format demonstrations as 
\hl{``[image] Describe the image in English in one sentence: [caption]''};
For VQA we format demonstrations as \hl{``[image]$\backslash$n[question] Answer in a word: [answer]''};
For Hateful Memes, we prompt the model with \hl{``[image] is an image with [text] written on it. Is it hateful? Answer: [answer]''}.
Notably, following the previous studies~\cite{alayrac2022flamingo,awadalla2023openflamingo}, we explicitly provide OCR text as inputs of AIM and baselines, and we don't extra prompt AIM can respond ``unanswerable'' in Vizwiz, reducing inappropriate induction.

\subsection{Baselines}
Considering our aim to enable any MLLMs to embrace efficient ICL, the underlying backbones within AIM are convinced baselines to compare their efficiency and performance, namely, QWen-VL and LLaVA-Next.
We also cite the results of the comparative MLLMs from their published studies for reference:
\begin{itemize}
    \item Flamingo: Similar to AIM, Flamingo, with the inner LLM having 7B parameters, breaks the quality between visual and textual modality and achieves outstanding performance on large-shot settings.
    \item OpenFlamingo: OpenFlamingo reproduces Flamingo in image-text tasks on the public dataset MMC4 which is also involved in the training set of AIM, with the inner LLM having 7B parameters.
\end{itemize}
Other MLLMs focus on multimodal ICL such as Otter~\cite{li2023mimic} and MMICL~\cite{zhao2023mmicl} due to \textbf{non-overlapped evaluating datasets} (Otter) and \textbf{different model sizes} (MMICL).

\subsection{Result}
We filter the benchmarks occurring in the training of our selected backbones to simulate the practical in-context learning situation.
Interestingly, when provided QWen-VL with demonstrations including textual information merely (\textit{w/o visual} in Table~\ref{tab:results}), it even outperforms the large shots situation provided both visual and textual.
Additionally, QWen-VL produces significant performance degradation in all 4 benchmarks when provided with over 8 demonstrations.
This further highlights that treating visual and textual tokens equally limits MLLMs from exhibiting outstanding ICL performance, despite QWen-VL having developed multi-image understanding ability during training.
When concentrating on LLaVA-Next, especially in the close-ended evaluation, perplexities concerning golden labels become \textbf{NotaNumber (NaN)} in 8- and 16-shot settings, occurring overflow while calculating PPL.
In other vision language tasks, LLaVA-Next fails to generate when provided over one demonstration and occurs <1 metric in evaluation since it didn't learn to understand interleaved image-text prompts during pre-training.

AIM aggregates image information into its linguistic space before generation.
The input format is close to prompts containing a single query image, significantly bridging the modal gap.
In this case, MLLMs are only required to attend to the query image while fused tokens still guide generation, thus bringing more concise responses.
Additionally, the valuable merit artfully unlocks the ICL ability of MLLM trained on the single image-text pair.
It is verified by the successful deployment on LLaVA-Next that the fused tokens combined with image and text information are harmless for the inner Vicuna.

For vision language tasks involved in this spectrum, QWen-based AIM outperforms backbones provided with concrete visual features that achieve +18 CIDEr gains in 16-shot in Flickr30k.
In the Vizwiz dataset, over 33\% answers are answerable in the statistic.
AIM exhibits relatively lower performance compared with other multimodal ICL methods since we don't prompt AIM to output `unanswerable', avoiding not solid short-cut answers.

Both OpenFlamingo and AIM employ MMC4 as the multi-image training set, but AIM, even applying LLaVA-Next as the backbone, still achieves comparable or even outstanding performance via aggregating image information when provided with random demonstrations (refer to the \textit{Random} row).
(Open) Flamingo applies \textbf{RICES} (Retrieval-based In-Context Example Selection)~\cite{yang2022empirical} to select demonstrations in latent space and achieve better performance.
The relevant results of AIM applying \textbf{RICES} are provided in Table~\ref{tab:rices} to discuss performance variance concerning random and well-retrieved demonstrations.

\section{Analysis}
\subsection{Training Data Abalation}
To avoid the model learning to generate short-cut answers because of in-domain or task-relevant training data, we train the linear layer on MMC4, which is a popular image-text-interleaved pre-training dataset used in OpenFlamingo.
However, training AIM on MMC4 is still potential to help MLLMs better understand image-text-interleaved inputs, achieving performance gains.

Due to AIM modifies the input form of MLLMs, we adopt LoRA to tune the built-in MLLMs, the QWen-VL and the LLaVA-Next, with comparable trainable parameters (17M) as AIM in Table~\ref{tab:results} by setting the LoRA rank to 32 to simulate these gains(\textbf{referring to $\star$ in Table~\ref{tab:results}}).
Further tuning on MMC4 improves LLaVA-Next in reading ICL prompts to some extent, but LLaVA-Next still underperforms providing pure text demonstrations (\textit{w/o visual}).
Due to QWen-VL having developed the multi-image understanding ability during pre-training, further training on MMC4 is not necessary, exhibiting even poorer performances since the web corpora is quite different from captioning, VQA, or image understanding.

\subsection{LLM Layer Count for Aggregation}
Considering the first layers directly interact with pre-trained embeddings, we perform ablation experiments on the first half (\textit{16}), and the first three-quarters (\textit{-24}) to explore the number of LLM layers required to aggregate image information in Table~\ref{tab:results}.
It is interesting that QWen-VL prefers the first 16 layers, while LLaVA-Next is inclined to full layers being the aggregator.
Therefore, the label words claim~\cite {wang2023label} that shallow layers (first half) focus on information aggregation is not completely applicable for LLaVA-Next in multimodal settings.

From a posterior view, LLaVA-based AIM obtains stable performance gains with the aggregating layers become deeper.
We attribute this conflict to LLaVA being pre-trained on single images, requiring deeper LLM layers to fuse image information into corresponding label space thoroughly, thus reducing the understanding difficulty for the built-in LLM.
Additionally, The Connecter in the LLaVA-Next project visual features from ViT to 576 visual tokens for a 336*336 image, while QWen-VL has 256 learnable queries.
Therefore, LLaVA-based AIM requires deeper LLM layers to perform visual information gathering.

\subsection{Perplexity Tendency of ICL}
\begin{figure}
    \centering
    \includegraphics[width=1\linewidth]{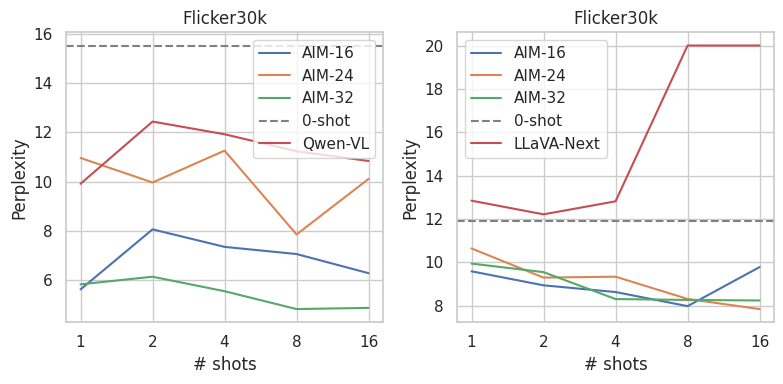}
    \caption{The perplexity variation tendency corresponds to the number of demonstrations. 0-shot server as the baseline.}
    \label{fig:perplexity}
\end{figure}

We briefly illustrate the perplexity variation tendency concerning golen labels of Flickr30k in Figure~\ref{fig:perplexity} with the number of demonstrations changing from 0 to 16.
Notably, the perplexity blast occurring in both QWen-VL and LLaVA-Next in the large shot setting indicates that the provided demonstration sequences significantly confused the underlying backbones, resulting in bad responses.
While in the scope of AIM, the perplexity presents a decreasing tendency in general with some noise brought by randomly sampled demonstrations.
Additionally, the most perplexity values are inferior to the 0-shot ones, indicating the provided demonstrations have a positive effect on helping MLLM generate current golden label responses.
\begin{figure}[t]
    \centering
    \includegraphics[width=\linewidth]{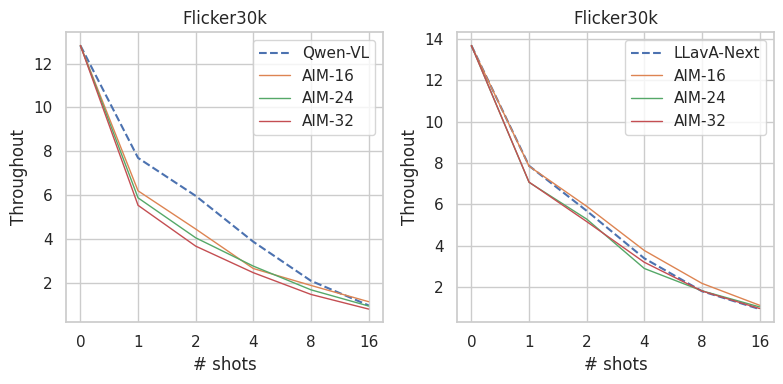}
    \caption{The throughput (iter/s) variation tendency is evaluated on a single H800, with the number of demonstrations increasing from 0 to 16.}
    \label{fig:throughput}
\end{figure}

\subsection{Inference Throughput of AIM}
AIM utilizes the inner LLM of existing MLLMs to complete image information aggregation operation.
This character makes AIM not need to load the other ``aggregator'', alleviating the memory costs.
However, image information aggregation requires inevitable but minimal time costs due to the parallel computation of forward propagation.
What's more, AIM drops all the visual tokens after aggregating them into the dense label space, which compensates for aggregation time costs to some extent by reducing the input length during generation\footnote{For QWen-VL, AIM drops 256 visual tokens consistently, and AIM drops 576 visual tokens for a 336*336 resolution image in LLaVA-Next.}.
We evaluate the throughput (iter/s) of AIM on Flick30k in Figure~\ref{fig:throughput}.

Overall, in the few shot settings (less than 8), naive MLLMs are more efficient than AIM, but AIM has a lower inference latency increment.
AIM becomes more efficient than the underlying backbone when provided with over 16 demonstrations.


\begin{table}
  \centering
    \begin{adjustbox}{width={\linewidth},totalheight={\textheight},keepaspectratio}
    \begin{tabular}{l|c|cccc|c}
    \bottomrule
    \multirow{2}{*}{Method} &  \multirow{2}{*}{\# Visual Tokens}& \multicolumn{4}{c|}{Avg. Textual Tokens}  & \multirow{2}{*}{Avg. Ratio} \\
    \cline{3-6}
    & & Flickr30k & OKVQA & Vizwiz & Hateful Memes \\
    \hline
    AIM-QWen & 256 & 30.1 & 16.0 & 15.7& 29.3 & 8\% \\
    \hline
    AIM-LLaVA & 576 & 34.4 & 18.3  & 17.9& 33.6 & 4\% \\
    \toprule
    \end{tabular}%
\end{adjustbox}
    \caption{Quantity statistics of visual and textual tokens in multimodal demonstrations.}
  \label{tab:ratio}%
\end{table}%
\subsection{Memory Cost of AIM}
The normal attention mechanism is known as $O(W^2)$ space complexity concerning a sequence with $W$ words.
Therefore, the length challenge brought by in-context demonstrations stimulates memory explosion straightforwardly.
AIM drops the visual tokens after image information aggregation and the remaining ratios of fused tokens $\mathcal{R}$ can be calculated according to the number of visual and textual tokens, denoted as $|V|$ and $|T|$:
\begin{equation}
    \mathcal{R} = \frac{|T|}{|V| + |T|}.
\end{equation}
We demonstrate the quantity statistics over four datasets in Table~\ref{tab:ratio}, indicating that LLaVA-based AIM merely retains about 4\% origin tokens in each multimodal demonstration.
Although LLaVA-Next integrates FlashAttention, dropping visual tokens still saves noticeable memory costs as illustrated in Figure~\ref{fig:memory}.
Notably, even if vision-language tasks have extremely long textual labels in assumption, AIM is capable of performing efficient ICL as normal with $\mathcal{R}$ close to 1 since visual tokens have been dropped and the textual tokens are required anyway.

\section{Conclusion}
In this paper, our initial exploration delves into the attention distribution within the multimodal ICL, revealing that the MLLM exhibits a greater emphasis on the linguistic modality.
Built upon this discovery, we propose a light multimodal framework AIM aiming to let any MLLMs embrace efficient ICL, which aggregates the image information of demonstrations into their dense latent space of demonstrated texts.
Generally, AIM transforms the multimodal ICL demonstration sequence into a form resembling a single query image accompanied by textual tokens.
Thereby, AIM successfully coordinates with any MLLMs regardless of their initial support for multimodal ICL.
Except for the outstanding performance of AIM compared with MMLMs specifically trained on multimodal ICL, AIM is both training and inferencing efficient due to its frozen de facto backbone and dropping hundreds of visual tokens.

\bibliographystyle{acl_natbib}
\bibliography{tacl2021}

\appendix
\section{Limitation and Discussion}
Although AIM makes MLLMs embrace efficient ICL regardless of their backbone support reading multi-images initially, there are still limitations.
The token remaining ratio in the original demonstration is determined according to the number of textual tokens, and therefore AIM will obtain less minimal efficiency gains if the labels are extremely over-length, despite the visual tokens being dropped and the textual tokens will exist anyway.
Additionally, caching all demonstrations takes up a lot of storage space with the demonstrations increasing as well, and we are attempting to quantify the virtual demonstrations to alleviate this problem.

\section{Evaluation Using RICES}
We mainly discuss using randomly sampled in-context demonstrations in the main text.
We employ RICES to retrieve similar demonstrations by measuring the image cosine similarity according to CLIP ViT-L/14.
The relevent results are presented in Table~\ref{tab:rices}.

\begin{table*}
\centering
\begin{adjustbox}{width={\textwidth},totalheight={\textheight},keepaspectratio}
\begin{tabular}{l|c|ccccc|ccccc|ccccc|ccccc}
\bottomrule
\multirow{4}{*}{Method} & \multirow{4}{*}{LLM}  & \multicolumn{5}{c|}{Flickr30k}  & \multicolumn{5}{c|}{OKVQA} & \multicolumn{5}{c|}{Vizwiz} & \multicolumn{5}{c}{Hateful Memes}\\
& & \multicolumn{5}{c|}{\textit{\small CIDEr $\uparrow$ }}& \multicolumn{5}{c|}{\textit{\small VQA-ACC $\uparrow$}}& \multicolumn{5}{c|}{\textit{\small VQA-ACC $\uparrow$}}& \multicolumn{5}{c}{\textit{\small ACC. $\uparrow$}}\\
\cline{3-22}
& & \multicolumn{5}{c|}{\textit{\small\#-shots}}& \multicolumn{5}{c|}{\textit{\small\#-shots}}& \multicolumn{5}{c|}{\textit{\small\#-shots}}& \multicolumn{5}{c|}{\textit{\small\#-shots}}\\
&&  1 & 2  & 4& 8& 16 &  1 & 2  & 4& 8& 16 &   1 & 2  & 4& 8& 16 &   1 & 2  & 4& 8& 16 \\
\hline
\textit{QWen-VL} & \multirow{4}{*}{QWen (7B)} 
&76.5&79.3& 78.4&72.9&70.4 
&47.3&57.6&56.2& 54.3& 53.5 
&30.2 &31.4&32.5& 32.2& 30.0 
&61.7&62.4&64.0& 62.9& 59.2  \\

\textit{AIM} &  & 
69.8&78.6&79.5&80.8&83.3 &
55.6&55.4&55.8&56.1&56.0 &
36.3&37.1&37.9&38.7 & 39.4 &
63.4&65.2&65.4 & 66.1 & 65.7 \\

\quad\quad\textit{-16}  & &
60.3&{81.3}&{80.9}&{84.2}&{85.6}&
{58.0}&{59.6}&{59.3}& {60.1}&{61.2} &
32.1&{37.2}&{38.5}&{39.1} & {40.2}&
64.8&66.7&{67.1}& {66.9}& {67.3}  \\

\quad\quad\textit{-24} &  &
64.2&79.2&80.7&82.6&84.5 & 

57.3&58.1&59.4&60.8& 61.2 & 
33.2&36.2&36.8&39.1& 38.9 &
58.6&59.6&61.2& 63.6& 64.0  \\
\hline

LLaVA-Next & \multirow{4}{*}{Vicuna (7B)} &
<1 & <1& <1& <1& <1 & 
<1 & <1& <1& <1& <1 & 
16.9&13.1&<1& <1& <1
&53.8&53.3&54.4& NaN& NaN  \\

\textit{AIM} & & 
53.9&55.7&60.3&{62.1}&{64.6}&
{57.3}&{59.2}&{62.8}&{61.9}&{62.3}& 
25.8& 26.9&{28.3}&{28.9}&{31.2}&
60.6&{61.4}&{62.0}&{63.4}&{65.1}\\
\quad\quad\textit{-16} &    & 
52.1&53.8&46.5&34.1&28.6&
53.1&52.4&53.1&54.7&51.6 & 
23.2&24.6&22.8&19.7&18.4&
59.3&60.1&51.5&52.1&53.9 \\
\quad\quad\textit{-24} &   & 
53.8&56.2&54.3&51.4&47.1&
55.6&56.7&57.1&{59.2}&54.1& 
{24.6}&{25.1}&26.3&25.5&23.3&
{59.7}&61.2&62.4&58.7&55.9\\

\toprule
\end{tabular}
\end{adjustbox}
\caption{Evlaution using RICES.}
\label{tab:rices}
\end{table*}

\section{Information Flow}
\begin{figure}[h]
    \centering
    \includegraphics[width=.5\textwidth]{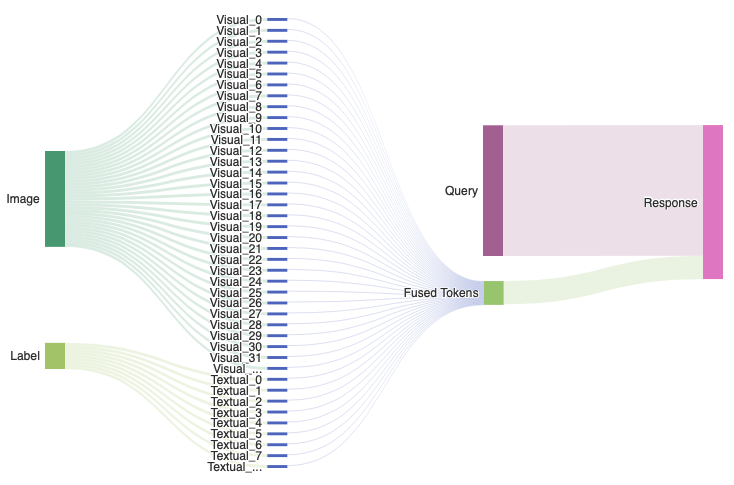}
    \caption{The information flow of How AIM performs image information aggregation (left) and response generation (right).}
    \label{fig:flow}
\end{figure}

To better explain how AIM performs image information aggregation and response generation, we illustrate the information flow of AIM in Figure~\ref{fig:flow}.
Specifically, the image and labels derivate visual tokens and textual tokens, respectively, where the textual ones are much shorter than the visual ones.
Then, AIM aggregates the visual information into dense latent space of labels through forward propagation that gets the last hidden states on top of the labels and converts them into fused tokens.
Finally, fused tokens, serving as in-context demonstrations, together with the current query, are fed into the inner MLLM for response generation.
\section{Perplexity and Throughput Analysises}
\begin{figure*}[h]
    \centering
    \includegraphics[width=.85\textwidth]{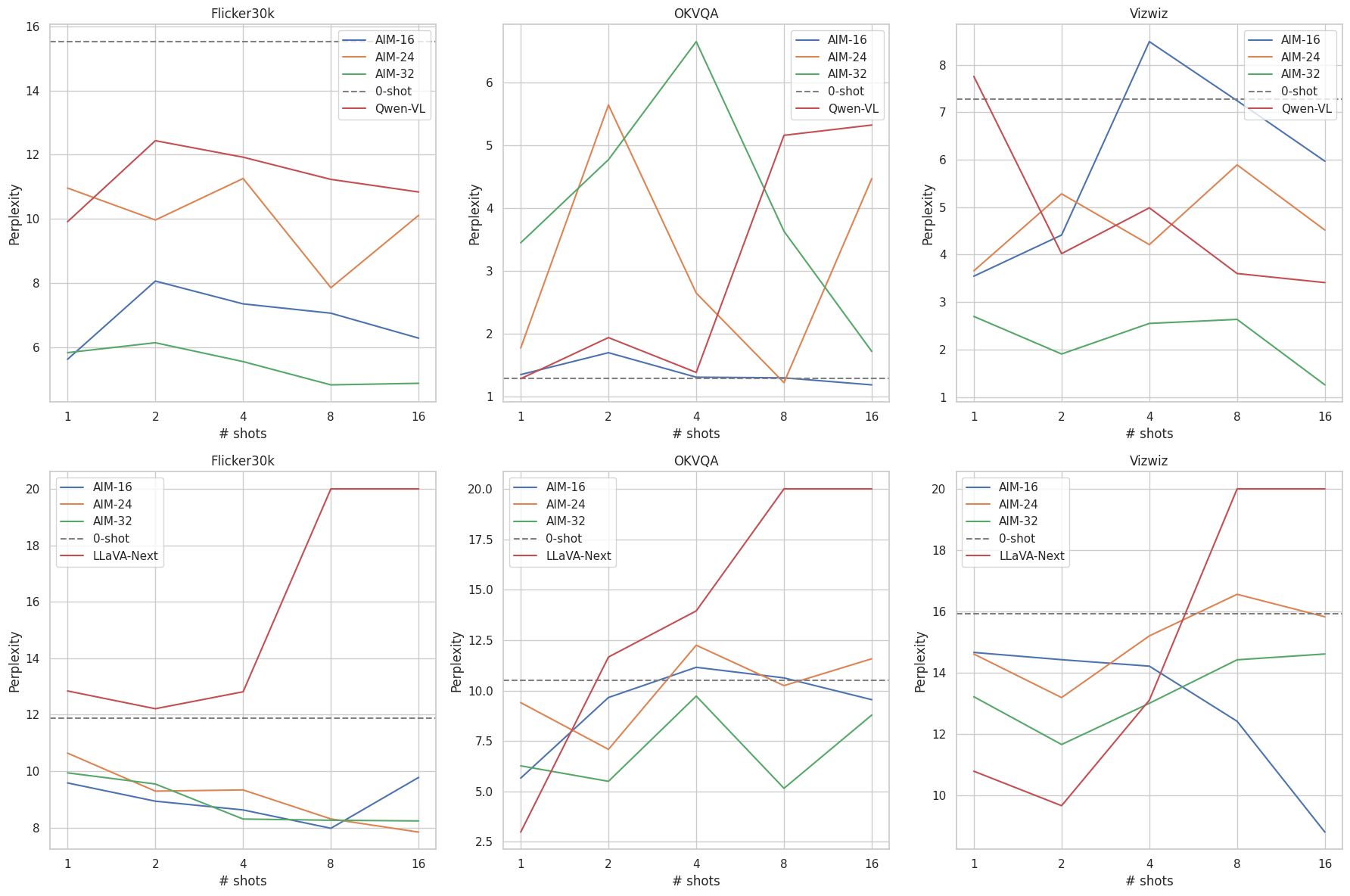}
    \caption{Perplexity analysis on Flickr30k, OKVQA and Vizwiz, with the backbone of QWen-VL and LLaVA-Next.}
    \label{fig:ppl}
\end{figure*}

Since Hateful Memes utilizes close-ended evaluation reflecting PPL implicitly, we provide the perplexity tendency on Flickr30k, OKVQA, and VizWiz in Figure~\ref{fig:ppl}, with the backbones of QWen-VL and LLaVA-Next.
Additionally, we demonstrate the throughput variation with respect to the number of demonstrations on Flickr30k, OKVQA, Vizwiz, and Hateful Memes in~\ref{fig:throughput}, with the backbones of QWen-VL and LLaVA-Next.
\section{Training Loss Curve}
\begin{figure}[h]
    \centering
    \includegraphics[width=.5\textwidth]{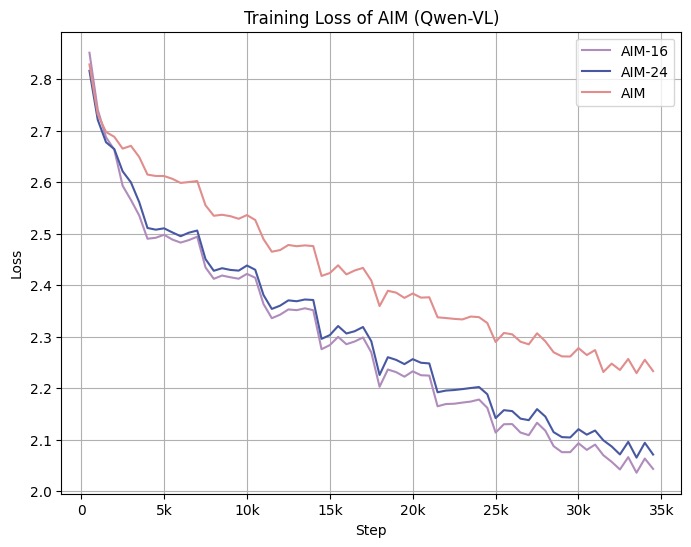}
    \caption{Training loss curve of AIM, with QWen-VL being the backbone.}
    \label{fig:lossqwen}
\end{figure}

\begin{figure}[h]
    \centering
    \includegraphics[width=.5\textwidth]{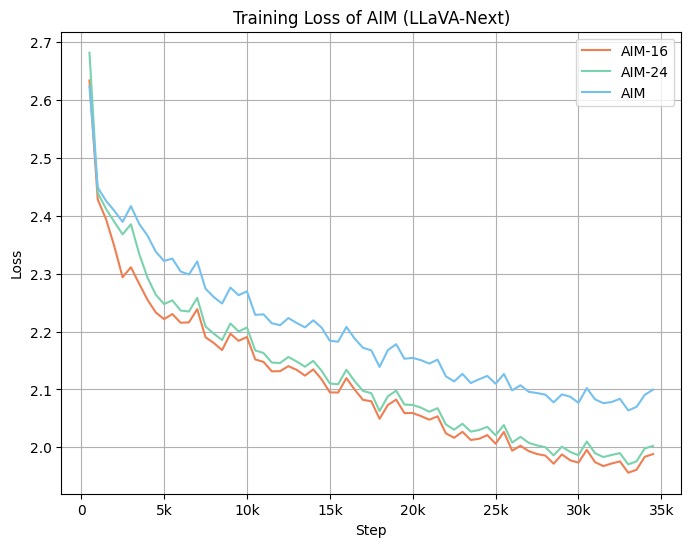}
    \caption{Training loss curve of AIM, with LLaVA-Next being the backbone.}
    \label{fig:lossllava}
\end{figure}
We utilize Huggingface Trainer to optimize AIM for about 12 GPU hours on a single H800 node, with the default optimizer and scheduler.
We demonstrate the training loss of AIM per 500 steps in Figure~\ref{fig:lossqwen} and Figure~\ref{fig:lossllava}.

\begin{figure*}
    \centering
    \includegraphics[width=.85\textwidth]{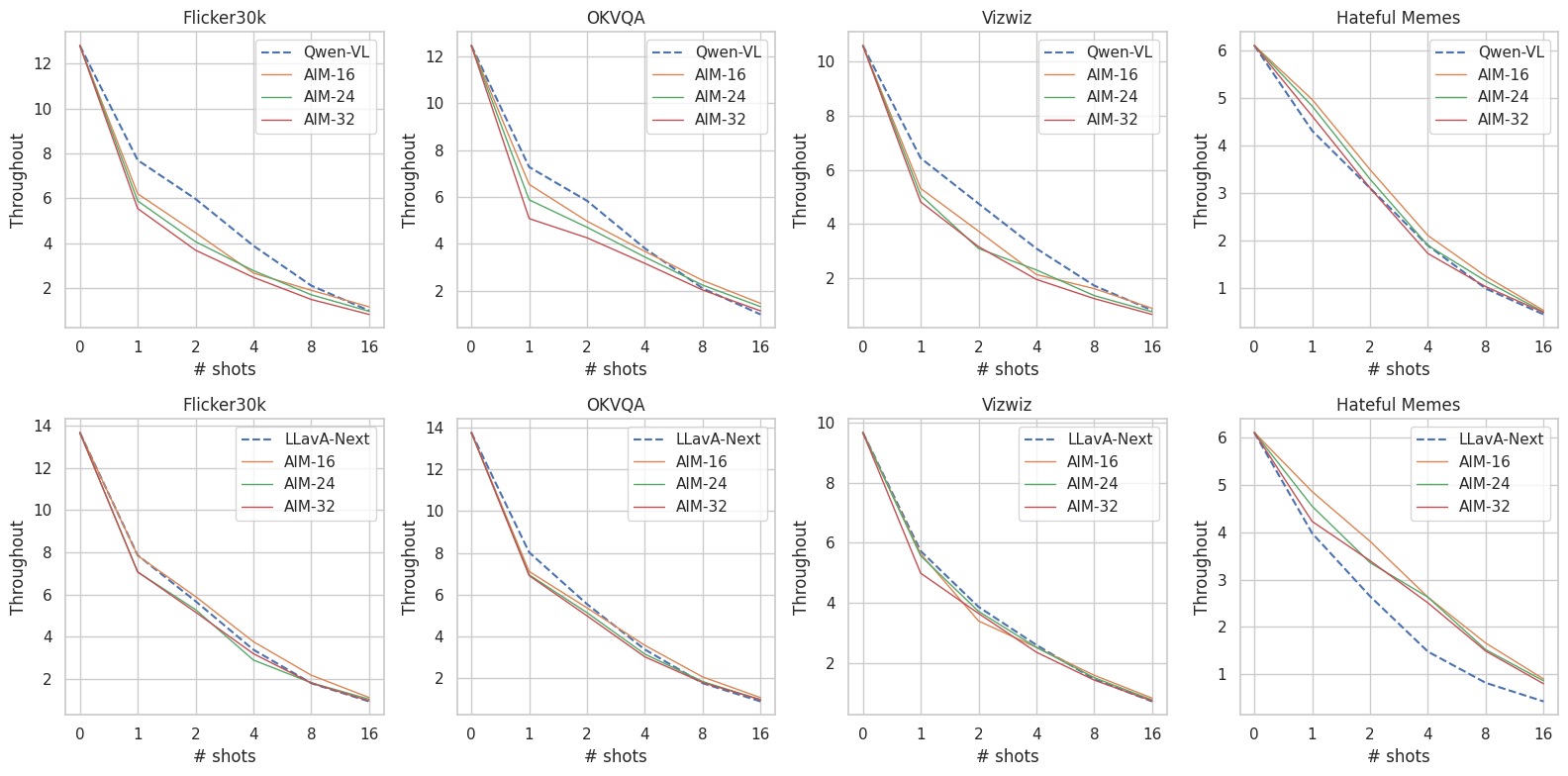}
    \caption{Throughput analysis on Flickr30k, OKVQA, Vizwiz and Hateful Memes, with the backbone of QWen-VL and LLaVA-Next.}
    \label{fig:throughput}
\end{figure*}

\end{document}